\documentclass[a4paper,10pt,prl,twocolumn]{revtex4-1}
\usepackage{color,amsmath,amssymb,amsthm,times,graphics,graphicx,bm,bbm,dcolumn}

\newcommand{\be}{\begin{equation}}
\newcommand{\ee}{\end{equation}}

\def\tr{ {\rm{Tr }}}

\begin{document}
\title{Noise-enhanced classical and quantum capacities in communication networks}

\author{Filippo Caruso}
\affiliation{Institut f\"{u}r Theoretische Physik, Albert-Einstein-Allee 11, Universit\"{a}t Ulm, D-89069 Ulm, Germany}

\author{Susana F. Huelga}
\affiliation{Institut f\"{u}r Theoretische Physik, Albert-Einstein-Allee 11, Universit\"{a}t Ulm, D-89069 Ulm, Germany}

\author{Martin B. Plenio}
\affiliation{Institut f\"{u}r Theoretische Physik, Albert-Einstein-Allee 11, Universit\"{a}t Ulm, D-89069 Ulm, Germany}
\begin{abstract}
The unavoidable presence of noise is thought to be one of the major problems to solve in order to pave the way for implementing quantum information technologies in realistic physical platforms.
However, here we show a clear example in which noise, in terms of dephasing, may enhance the capability of transmitting not only classical but also quantum information, encoded in quantum systems, through communication networks. In particular, we find analytically and numerically the quantum and classical capacities for a large family of quantum channels and show that these information transmission rates can be strongly enhanced by introducing dephasing noise in the complex network dynamics.
\end{abstract}
\maketitle
\paragraph{Introduction.--}
An important obstacle for the development of quantum communication
technologies is the difficulty of transmitting quantum information
over noisy quantum communication channels, recovering and refreshing
it at the receiver side, and then storing it in a reliable quantum
memory \cite{nc,bs}. This concerns both point-to-point communication
as well as more complex quantum networks consisting of several nodes.
These operations are necessary as the unavoidable
presence of noise during transmission via a quantum channel and its
processing at the receiver's end is generally expected to degrade the transmission
quality. It was pointed out however, that noise may in fact have a
positive influence on sustaining quantum correlations \cite{HuelgaPlenio}.
This motivated some early explorations of the potentially beneficial
effects that noise may have on information transmission through quantum
channels \cite{FPTH}. It was not possible however to compute capacities
in those examples and attention focussed on related quantities that
were furthermore restricted to certain input states. As a result, no firm conclusion
could be drawn from these considerations. Recently though, it was realised
in a different context that noise may have a positive impact on transport
phenomena in complex networks. In fact, it was found that excitation energy
transfer (EET) in light harvesting complexes during photosynthesis can
benefit considerably from the presence of dephasing noise
\cite{Aspuru,PlenioH08}. Here, it is the intricate interplay of
noise and quantum coherence that explains the remarkable efficiency,
well above $90\%$, for EET in light
harvesting complexes during photosynthesis whereas noise-free
systems exhibit efficiencies of around $50\%$ only \cite{ccdhp09}.
Motivated by these results, we study the scenario of a realistic
communication network, subjected to a noisy evolution, and derive
analytically and numerically the channel capacities. We will
demonstrate that both classical and quantum channel capacities
can increase thanks to the presence of dephasing noise. Remarkably, in the
case of quantum capacities, the presence of noise may lead to a
finite quantum capacity where the noiseless system has vanishing
capacity.
\begin{figure}[t]
\centerline{\includegraphics[width=.3\textwidth]{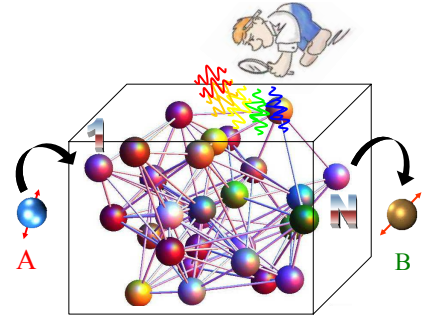}}
\caption{Communication network: on one side, Alice sends a message encoded in the qubit $1$, while the rest of the network is prepared in its ground state. On the other side, after some noisy evolution of the network state, Bob tries to recover Alice's message decoding his output qubit state in the site $N$.}\label{fig1}
\end{figure}
\paragraph{The Model.--}
We consider a generic complex network of $N$ vertices, in which each site
represents a two-level quantum system (qubit). Suppose that the sender of
the message, Alice (A), wants to transmit a message to the receiver, Bob
(B), by using such quantum network. The communication protocol can be the
following: i) Alice applies a swap operation in order to set the unknown
initial state $\rho_A$ in the site $1$, while the rest of the network is
initially prepared in the ground state $|0 \dots 0 \rangle$, ii) then they
let the network state evolve under some quantum noisy evolution, iii) at
time $t_{out}$, Bob tries to recover the information sent by Alice through
some decoding procedures applied to his output state $\rho_B(t_{out})$,
which corresponds to the reduced density operator for the $N$th qubit (up
to a local unitary transformation). Mathematically, at each time $t_{out}$,
one can describe this process as a completely positive and
trace-preserving (CPTP) quantum channel of the form
 \be
\rho_B(t_{out}) \equiv {\cal E}(\rho_A)=\tr_{E}[ U(t_{out}) (\rho_A \otimes \rho_E ) U(t_{out})^\dagger]
 \ee
where $\rho_E$ is the initial state of some environment and $U(t_{out})$
is the unitary evolution of system + environment for a time $t_{out}$.
For instance, in the case of an Hamiltonian evolution of the network,
$\rho_E$ is the ground state of all the qubits $2,\dots,N$ and
$U(t_{out})=\exp[-i H t_{out} /\hbar]$ is the unitary evolution operator
associated to the Hamiltonian $H$. In this case, following Ref. \cite{GF},
if the Hamiltonian commutes with the Pauli operator $\sigma_z$, i.e. the
number of qubits in the state $|1\rangle$ is constant in time, at each time
step $t_{out}$ the corresponding quantum map is an amplitude damping channel
${\cal D}(\eta)$, with the damping coefficient given by $\eta=\langle N|
U(t_{out}) |1\rangle$, with the convention that $|j\rangle$ denotes the
state in which all the qubits are in the state $\mid\downarrow\rangle$, except
the qubit $j$ in the state $\mid\uparrow\rangle$. In particular, Ref.
\cite{GF} considered a spin chain subjected to an Heisenberg Hamiltonian
evolution and in this context derived the channel capacities. Here, we
investigate a noisy evolution of a network of $N$ qubits, in which, for
instance, some pure dephasing noise is present in the dynamics and, as shown
later, will play a key role in the information transfer rates of the
corresponding communication channel. For simplicity, we will consider the
following Hamiltonian, $H = \sum_{j=1}^N \hbar\omega_j \sigma_j^{+}\sigma_j^{-}
        + \sum_{j\neq l} \hbar v_{j,l} (\sigma_j^{-}
        \sigma_{l}^{+} + \sigma_j^{+}\sigma_{l}^{-})$,
and a local Lindblad term that takes into account the dephasing caused by some
 surrounding environment, i.e.,  ${\cal L}_{deph}(\rho) = \sum_{j=1}^{N} \gamma_j[
        -\{\sigma_j^{+}\sigma_j^{-},\rho\} +
        2 \sigma_j^{+}\sigma_j^{-}\rho \sigma_j^{+}\sigma_j^{-}]$,
with $\sigma_j^{+}$ ($\sigma_j^{-}$) being the raising and
lowering operators for site $j$, $\hbar\omega_j$ being the local
site excitation energy, $v_{k,l}$ denoting the hopping rate of
an excitation between the sites $k$ and $l$, and $\gamma_j$ being
the dephasing rate at the site $j$. Let us point out, however, that
the following results must be valid for any Hamiltonian, provided
that it commutes with the operator $\sigma_z$, i.e. preserving the
number of `excitations' in the network evolution, and quite reasonably
for other forms of noise, also beyond the Markovian approximation.
Generalizing Ref. \cite{GF}, it is possible to show that,
at each time $t_{out}$, the corresponding CPTP quantum channel is of
the following form:
\be
{\cal E}(\eta,s): \ \rho_{A} = \left(
    \begin{array}{cc}
    {p} & \gamma\\
    \gamma^{*} & {1-p}
    \end{array} \right) \rightarrow \rho_B = \left(
    \begin{array}{cc}
    {\eta p} & \sqrt{\eta s}\gamma\\
    \sqrt{\eta s} \gamma^{*} & {1-\eta p}\end{array} \right)
    \label{map}
\ee
with $p$ a real number in the range $[0,1]$ and $\gamma$ a complex number such that $|\gamma|^2 \leq p(1-p)$. The channel is completely defined by two time-dependent parameters: i) $\eta \in [0,1]$ describing the population damping, and ii) $s\in
[0,1]$ includes the decoherence effects. On one hand, because of the linearity of a quantum channel, $\eta(t)$ corresponds to the population of site $N$ at time $t$ when $\rho_A=|1\rangle \langle 1 |$, i.e. one excitation is initially in the site $1$. On the other hand, $s$ is generally a complicate time-dependent function of all the parameters involved in the noisy evolution and later will be determined numerically considering a generic input qubit state. Note also that $s$ can be considered a real number because any phase $e^{i \theta}$ can be eliminated applying a local unitary transformation, i.e. $|1\rangle \rightarrow e^{-i \theta} |1\rangle$, and the quantities analyzed later are invariant under such operations. Besides, it can be easily shown that the map in (\ref{map}) is equivalent to a consecutive application of an amplitude damping channel ${\cal D}(\eta)$ and a phase-flip channel ${\cal N}(s)$ (unitarily equivalent to a dephasing channel), changing the phase of the state $|1\rangle$, i.e. $|1\rangle \rightarrow - |1\rangle$, with probability
$(1-\sqrt{s})/2$. In other words, one has ${\cal E}(\eta,s) = {\cal D}(\eta) \circ {\cal N}(s)
 = {\cal N}(s) \circ {\cal D}(\eta)$.
In general, a CPTP quantum channel can be also represented in an elegant form
known as operator-sum (or Kraus) representation \cite{nc}, i.e.
${\cal E}(\rho_A) = \sum_k A_{k} \rho_A A_{k}^{\dagger}$, where the so-called
Kraus operators $A_k$ satisfy the condition $\sum_{k} A_{k}^{\dagger} A_{k} =
\openone$. In particular, these operators for the map in (\ref{map}) are given
by $A_1=\text{diag}(0,\sqrt{s \eta})$, $A_2=\text{anti-diag}(0,\sqrt{1-\eta})$,
and $A_3=\text{diag}(0,\sqrt{(1-s)\eta})$. In fact, one can show that
${\cal E}(\rho) =\sum_{k=1}^{3} A_{k} \rho_A A_{k}^{\dagger}$ corresponds
exactly to $\rho_B$
as in Eq. (\ref{map}).
In the following, we will analyze the capability of the channel in (\ref{map})
for transmitting classical and quantum information asymptotically undisturbed,
i.e. with a vanishing error probability in the limit of long messages and with
some optimal encoding and decoding schemes, by calculating, respectively, the
classical and quantum channel capacities and some related figures of merit.
\paragraph{Classical Capacity.--}
\begin{figure}[t]
\centerline{\includegraphics[width=.38\textwidth]{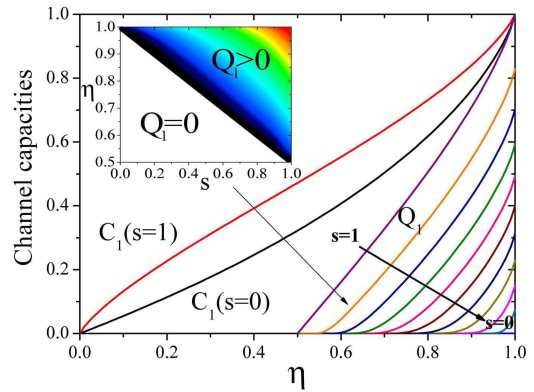}}
\caption{Classical and quantum capacities, $C_{1}({\cal E}(\eta,s))$ and
$Q_{1}({\cal E}(\eta,s))$, versus $\eta$ and $s$. Inset: contour plot for
$Q_{1}({\cal E}(\eta,s))$ as a function of $s$ and $\eta$ in a color gradient
representation where red corresponds to $1$ and black to vanishing values of
$Q_1$. The quantum capacity is always zero for $\eta \leq 1/2$ and any $s$.}
\label{fig2}
\end{figure}
The so-called `one-shot' or product state classical capacity of a quantum
channel is obtained maximizing the Holevo information \cite{holevo}, i.e.
\be
        C_{1}({\cal E}) = \max_{\xi_k,\rho_k} \left[ S\left(\sum_k \xi_k
        {\cal E}(\rho_k) \right) - \sum_k \xi_k S\left({\cal E}(\rho_k)
        \right)\right]
\ee
where the maximum is taken over all probability distributions $\xi_k$ and
collections of density operators $\rho_k$ and $S(\rho)=-\tr[\rho \log_2 \rho]$
is the von Neumann entropy of the state $\rho$. This is the classical
capacity  with unentangled encodings while a full maximization over multiple
channel uses would provide the unrestricted classical capacity of a quantum
channel. For simplicity, we will focus here on the case of unentangled
encodings which yields lower bounds for the unrestricted classical capacity
of a quantum channel. Notice, however, that the classical capacity with entangled
encodings is a monotonic increasing function of $\eta$ (by the bottleneck inequality),
and then, in the presence of a dephasing-induced enhancement of the population transfer ($\eta$), also the full classical capacity is enhanced.
Now, we calculate $C_1$ for the family of quantum channels in
(\ref{map}). In the case $s=1$, the channel reduces to ${\cal D}(\eta)$,
whose channel capacities were studied in Ref. \cite{GF}. Here, first we
will solve analytically the case of $s=0$, which will be relevant in the
example below. Let us consider an ensemble $\{\xi_k,\rho_k\}$, with $\rho_k$
of the form of $\rho_A$ in Eq. (\ref{map}), with parameters $p_k$ and
$\gamma_k$. The ensemble average state will be a state $\rho$ with
coefficients $p$ and $\gamma$ such that $p= \sum_k \xi_k p_k$ and
$\gamma= \sum_k \xi_k \gamma_k$. In the case of $s=0$, it can be easily
shown that the Holevo information reduces to $H_2(\eta p) - \sum_k \xi_k
H_2(\eta p_k)$, with $H_2$ being the binary entropy function defined as
$H_2(x)=-x \log_2 x -(1-x) \log_2 (1-x)$. By exploiting the concavity
property of $H_2$, one finds that $C_{1}({\cal E}(\eta,0)) \leq \max_{p
\in [0,1]} [H_2(\eta p) - p H_2(\eta)]$. The last step of this analytical
calculation consists of showing that this upper bound is actually tight and
can be obtained with the following optimal ensemble defined by the
coefficients $\xi_1=p$ and generic $\xi_i$ such that $\sum_{k} \xi_k=1$,
with $p_1=1$ and $p_i=0$, for $i=1,\dots d$, where $d$ is the generic
dimension of the ensemble $\{\xi_k,\rho_k\}$. Therefore, we find that
\be
\label{class_cap}
C_{1}({\cal E}(\eta,0)) = H_2(\eta \bar{p}) - \bar{p} H_2(\eta) \; ,
\ee
where $\bar{p}=[(1-\eta)^{(\eta-1)/\eta}+\eta]^{-1}$ is the optimal value
of $p$, which we find analyzing that function analytically -- see Fig. \ref{fig2}.
Notice that,
because of the composition law above, one has also that
${\cal E}(\eta,s_1) = {\cal E}(\eta,s_2) \circ {\cal N}(s_1/s_2)$ for
$s_1 < s_2$. Hence, by the bottleneck inequality, the following relations
hold, i.e. $C_{1}({\cal E}(\eta,0)) \leq C_{1}({\cal E}(\eta,s_1)) \leq
\dots \leq C_{1}({\cal E}(\eta,s_n)) \leq C_{1}({\cal E}(\eta,1)) \equiv
C_{1}({\cal D}(\eta))$ for $0 \leq s_1 \leq \dots \leq s_n \leq 1$.
In other words the $\eta$-dependence of
$C_{1}({\cal E}(\eta,s))$ will be described by a continuous family of
lines for intermediate value of $s$ in the range $[0,1]$, i.e. between
the red ($s=1$) and black ($s=0$) extreme lines in Fig. \ref{fig2}. From
these results it turns out that the presence of dephasing, on one hand,
may increase the value of $\eta$ enhancing also $C_1$, but, on the other
hand, increasing the amount of decoherence in the network (i.e., decreasing
$s$), may reduce the value of $C_1$. Hence, a compromise of
these two dephasing-induced effects could lead to a global enhancement of
the classical capacity, as shown below.
\paragraph{Example: FMO-channel capacity.--}
\begin{figure}[t]
\centerline{\includegraphics[width=.34\textwidth]{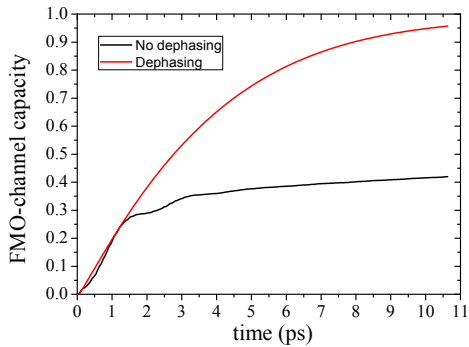}}
\caption{Classical capacity of the FMO-complex quantum channel,
$C_1({\cal E}(p_{sink}(t),0))$, vs. time, in the absence and presence of
dephasing. There is a clear remarkable
enhancement of $C_1$ in the case of pure dephasing noise.}\label{fig3}
\end{figure}
Here, we will describe a concrete example of the results above. In particular,
we will consider the transport dynamics of electronic excitations in a
biological pigment-protein complex, called FMO (Fenna-Matthews-Olson) complex,
involved in the early steps of photosynthesis in sulphur bacteria
\cite{fleming07b}. It is possible to describe this complex as a network of
$7$ chromophores or sites, represented as qubits. The coherent exchange of excitations
between sites can be described by the Hamiltonian above plus local Lindblad terms that take into account the dephasing and
dissipation caused by the surrounding environment -- see Ref. \cite{ccdhp09} for more details about this model. Actually, the channel in Eq. (\ref{map}) preserves the same form also in presence of dissipation, which will simply affect the values of $\eta$ and $s$. Moreover, the exciton energy is irreversibly transferred to the so-called reaction center, where it is immediately (irreversibly) converted into chemical energy. At each time $t$ (in a scale of order of $\mathrm{ps}$), this evolution can be mapped as in Eq. (\ref{map}), where $\eta$ is equal to the amount of excitations $p_{sink}(t)$ in the reaction center with one excitation initially in site $1$, while $s=0$ since there is just a population transfer from the site $3$ to the reaction center, i.e.  the map is ${\cal E}(p_{sink}(t),0)$.
By choosing the parameters as in Ref. \cite{ccdhp09}, we find that the classical capacity of the FMO complex dynamics, described as a quantum channel, $C_1({\cal E}(p_{sink}(t),0))$, is remarkably enhanced in the presence of dephasing,
especially after $1~\mathrm{ps}$ -- see Fig. \ref{fig3}. The dephasing enhanced classical capacity is due to the acceleration of transport in the network and may have been expected from the results of \cite{Aspuru,PlenioH08,ccdhp09} as dephasing noise does not affect classical information. Dephasing noise however destroys quantum information and it is therefore not immediately evident that the quantum capacity may be enhanced by dephasing as well.
\paragraph{Quantum capacity.--}
The quantum capacity $Q$ refers, instead, to the coherent transmission of
quantum information (measured in number of qubits), i.e. quantum states,
through a quantum channel. It is more difficult to treat than the classical
capacity discussed above and its explicit calculation
is one of the basic issues in quantum information science. The single-use
formula for the quantum capacity is obtained maximixing the coherent
information \cite{lds}, i.e.
\begin{eqnarray}
        Q_1({\cal E}) \,=\, \max_{\rho \in {\cal H}} [S({\cal E}(\rho))-S(\rho,{\cal E})]
\end{eqnarray}
where the maximization is performed over all qubit states in the input Hilbert space $\cal H$. Here, $S(\rho,{\cal E})$ is the exchange entropy of the channel \cite{nc}, describing the amount of information exchanged between the system and the environment after the noisy evolution, and is given by $S(\rho,{\cal E}) \equiv S(W)=-\tr[W \log_2 W]$ with $W_{ij}=\tr[A_i \rho A_j^\dagger]$. Note that $Q_1$ is usually a lower bound for $Q$ (which is maximized over many channel uses), since the coherent information is generally not additive. As discussed above,
for simplicity, we will optimize numerically the coherent information over all possible input states $\rho_A$ as in (\ref{map}).
It turns out numerically that the expression to maximize is always decreasing in $|\gamma|^2$ for $\eta \geq 1/2$ and then it achieves the maximum value for $\gamma=0$; the remaining optimization in $p$ has been performed numerically and the results are shown in Fig. \ref{fig2}. In the case of $s=1$, the channel reduces to ${\cal D}(\eta)$, for which the coherent information can be proved to be additive (i.e., since it is a degradable channel \cite{ds}), and the optimization over the channel uses is not necessary. The behaviour of the quantum capacity $Q$ as a function of $\eta$ for $s=1$ has been investigated in Ref. \cite{GF}. Here, we generalize those results in the presence of dephasing, i.e. $s < 1$, and find numerically the one-shot quantum capacity $Q_1$ as in Fig. \ref{fig2}. The additivity of the coherent information cannot be proved for $s<1$ but our results for $Q_1$ are of course a lower bound for the capacity $Q$ with entangled encodings. Although we have not computed the full quantum capacity rigorously and analytically, our lower bounds are sufficient to show the noise-assisted enhancement for the unrestricted quantum capacity as well. Indeed, because of the composition law ${\cal E}(\eta,s) = {\cal D}(\eta) \circ {\cal N}(s)$, the channel in (\ref{map}) has always a vanishing $Q$ for $\eta \leq 1/2$ and for any $s$, since in this regime ${\cal D}(\eta)$ has $Q=0$ (being an anti-degradable channel) \cite{GF,CG}. Similar enhancement is observed in other figures of merit, i.e. channel fidelity $F(R)$ and entropy $S(R)$~\cite{KW,roga}, which can be analytically derived for the map in (\ref{map}), i.e. $F(R)=1/4(1+\eta+2 \sqrt{\eta s})$ and $S(R)=-\sum_{i=1}^3 \lambda_i \log_2 \lambda_i$, with $\lambda_1=(1-\eta)/2$, $\lambda_{2,3}=1/4(1+\eta \pm \sqrt{4 \eta s + (1-\eta)^2})$.
A specific example of a 3-qubit network is shown in Fig. \ref{fig4}. The presence of dephasing `switches on' the channel capability of transmitting quantum information and the optimal rates can be very close to one (i.e., almost perfect state transfer), while $Q$ is exactly zero without dephasing. The intuitive reason for this behaviour is the fact that, in the noise-free case, quantum
information progresses along two possible paths, thus being split and not
arriving at the same time. This approximates a channel that splits the quantum
information which in turn has vanishing quantum capacity. For strong dephasing
the path via $E$ is blocked and direct transfer from $A$ to $B$ leads to the
arrival of all quantum information at $B$. Noise assisted channel capacities
may also be observed for larger networks and when all sites suffer
dephasing, but the effect is most pronounced for non-uniform distribution
of the noise. As final remark, notice that the dephasing can be induced by the presence of an eavesdropper, Eve, in the third site, in a quantum cryptographic scenario and, interestingly enough, it turns out that the eavesdropping operation is completely useless for Eve (the corresponding quantum capacity for the channel $A \rightarrow E$ remains exactly zero), but it does sensibly improve the Alice-Bob communication.
\begin{figure}[t]
\centerline{\includegraphics[width=.34\textwidth]{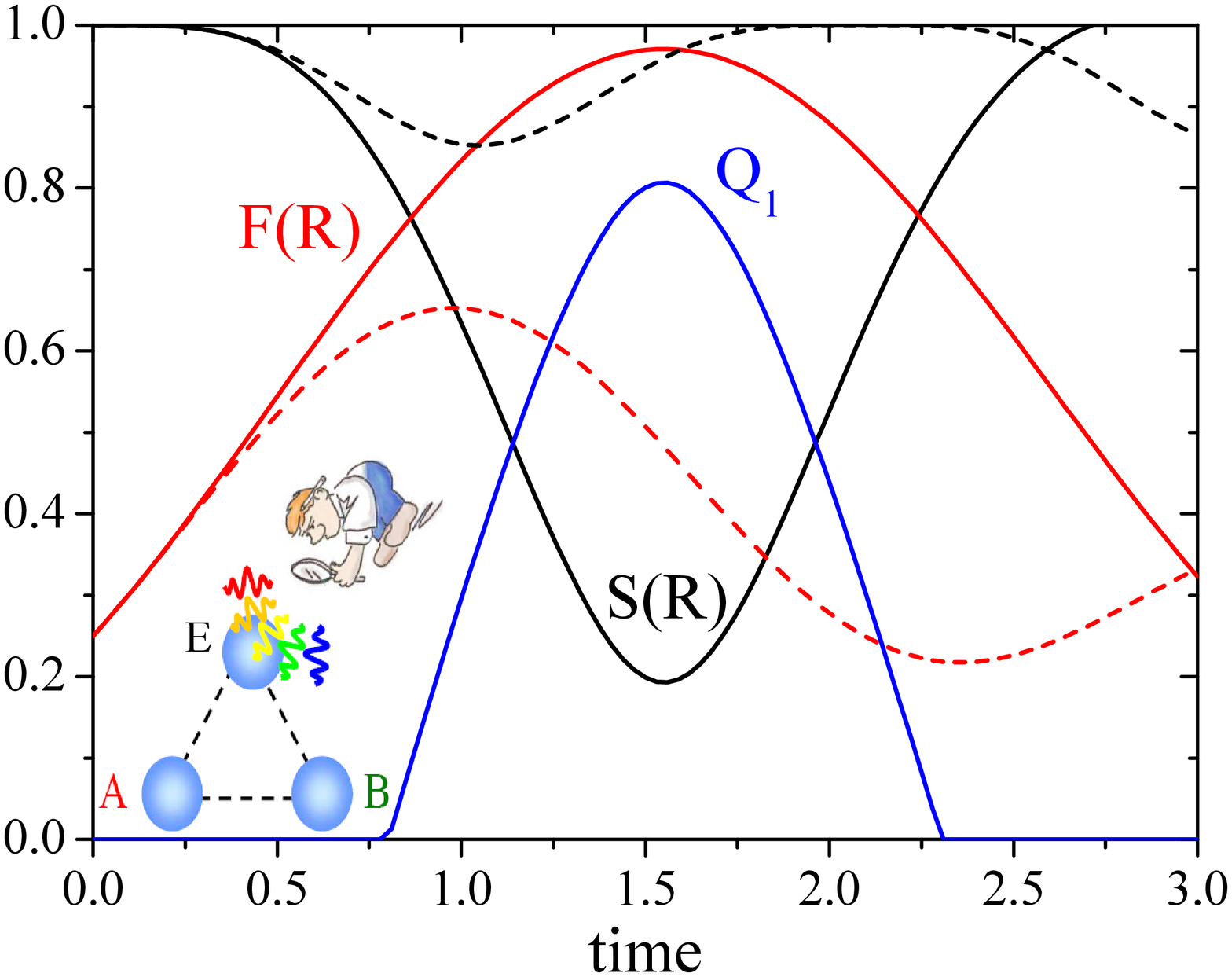}}
\caption{Quantum capacity $Q_1$ for a $3$-site network vs. time. In absence of noise, $Q$ is always zero, while $Q_1$ can get values almost up to one in the presence of dephasing. Here, the hopping and dephasing rates are $v_{12}=v_{23}=v_{13}=1$, and $\gamma_2=50$, $\gamma_1=\gamma_3=0$.
The corresponding channel entropy and fidelity are also shown in the noiseless (dashed) and dephasing (continuous line) case.}\label{fig4}
\end{figure}
\paragraph{Conclusions and Outlook.--}
We have evaluated analytically and numerically the classical and quantum channel capacities of a realistic communication network
and showed that these optimal information transmission rates can be enhanced by applying some pure dephasing to the network. In particular, this allows us to reinterpret the observed dephasing-assisted EET in
photosynthetic complexes as an example of a quantum channel with a noise-enhanced classical capacity. Perhaps more surprisingly, we have shown that the presence of noise may lead to a
finite quantum capacity where the noiseless system has vanishing
capacity. As a result, not only the transmission rate of classical information can be assisted by noise but also the transmission of quantum information coded in quantum states. The possible relevance of this result in the context of quantum cryptography has been illustrated in a simple scenario.
We expect these results to be easily generalizable to bosonic systems and to be valid for any Hamiltonian preserving the number of excitations, other forms of noise and also for non-Markovian evolutions. Finally, the three-site quantum network illustrating the fundamentals of our results could be experimentally investigated relatively easily by considering, for instance, quantum information platforms using trapped ions or cold atoms where forms of dephasing noise can be introduced in a controlled manner \cite{porras}.
\begin{acknowledgments}
This work was supported by EPSRC grant EP/C546237/1, the EU STREP
project CORNER, the EU Integrated project QAP, the Royal Society, a Marie Curie
Fellowship, and a Alexander von Humboldt Professorship.
\end{acknowledgments}
\end{document}